\documentclass[10pt,a4wide]{article}
\usepackage{newlfont}
\usepackage{amsmath}
\usepackage{graphics}
\usepackage{float}
\usepackage{graphicx}
\usepackage{epsfig}

\begin{document}

\title{Dynamical effects of QCD in $q^2 \bar{q}^{2}$ systems} \vspace{1cm}
\author{M. Imran Jamil \thanks{e mail: mimranjamil@hotmail.com},
Bilal Masud \thanks{e mail: bilalmasud@chep.pu.edu.pk}, \\
\textit{Centre For High Energy Physics, Punjab University,
Lahore-54590, Pakistan.}}
\date{}
\maketitle

\begin{abstract}
\vspace{.5cm}\noindent

We study the coupling of a tetraquark system to an exchanged meson-meson channel,
using a pure gluonic theory based four-quark potential {\em matrix} model which is
known to fit well a large number of data points for lattice simulations of different
geometries of a four-quark system. We find that if this minimal-area-based potential
matrix replaces the earlier used simple Gaussian form for the gluon field overlap factor $f$
in its off-diagonal terms, the resulting $T$-matrix and phase shifts develop an angle dependence
whose partial wave analysis reveals $D$ wave and higher angular momentum components in it.
In addition to the obvious implications of this result for the meson-meson scattering,
this new feature indicates the possibility of orbital excitations influencing properties
of meson-meson molecules through a polarization potential. We have used a formalism of
the resonating group method, treated kinetic energy and overlap matrices on model of
the potential matrix, but decoupled the resulting complicated integral equations
through the Born approximation. In this exploratory study we have used a quadratic
confinement and not included the spin-dependence; we also used the approximation of
equal constituent quark masses.
\end{abstract}

\section{Introduction}

Models of hadronic physics are to be compared with experimental
results as well as with our understanding of QCD for large momentum
transfers {\em and} for low momentum transfers where Feynman
diagrams are not useful. In this way, models may help us in improving
our knowledge of QCD for low and intermediate energies of interest
to hadronic spectroscopy and eventually nuclear physics. One way to
get this understanding is to note some features present in the
perturbative QCD, lattice gauge theory or models of atomic and
nuclear physics and check if these features can be used in the low
and intermediate energy hadronic physics.

One such feature is an approach based on pair-wise interaction for
an interacting multiparticle system (composed of more than two or
three particles). This has been successful in atomic and
many-nucleon systems; the corresponding two-body interaction being
described by Coulombic and Yukawa potential, for example. The
question is if the explicit presence of {\em Non-Abelian} gluon
field can also be replaced by {\em two body} interquark potentials.
The simplest way to use such a model is to try a {\em sum} of
two-body potentials or interactions, the usual approach of atomic
and nuclear physics. For comparison, it can be noted that the lowest
order perturbative Feynman diagrams amplitudes are of this form, and
a simple extension of this diagrammatic approach to multiquarks also
has this pattern; see ref. \cite{T. Barnes}, and the later ones in
the same approach, where the one gluon exchange potential, though,
is replaced by Coulombic-plus-linear-plus-hypefine. If the
numerically calculated energies of the four-quark systems on a
lattice, in the static quark limit, are compared with a model that
use only a {\em sum} of two-quark potentials, the model give a gross
overestimate of the (magnitude of) four-quark binding energies; see
fig. 4 of the same ref.\cite{Green1}. A gluon field overlap factor
$f$ was introduced \cite{B.Masud} essentially as a solution to this
discrepancy. This factor multiplied only off-diagonal elements of
the overlap, kinetic and potential energy matrices of the otherwise
pairwise-sum-based Hamiltonian in the three-dimensional basis of the
model system of four valence quarks/antiquarks and the purely
gluonic field between them. Initially \cite{B.Masud} the geometrical
dependence of $f$ on the quark positions was chosen purely based on
computational convenience and had no known comparison with any QCD
simulations. But when its different forms were compared \cite{P.
Pennanen} with two-colour lattice numerical simulations in the pure
gluonic theory, a factor of $\exp(-k S_{min})$ had to be included in
off-diagonal terms of the every version of the potential and overlap
matrices; $S_{min}$ is the minimal spatial area bounded by four
external lines joining the four quarks and $k$ is geometrically a
constant. Only this way, a version of the model was eventually able
to fit well "100 pieces of data---the ground and first excited
states of configurations from six (kinds of) different four-quark
geometries (including squares, rectangles, quadrilaterals and
tetrahedra) calculated on a $16^3\times 32$ lattice---with only four
independent parameters to describe the interactions connecting the
basis states". It is to be noted that this exponential dependence on
the spatial area in the model can be possibly traced back to the
related use of the space-time area in more familiar models of Wilson
loops studying time evolutions of a quark-antiquark pair. The
connection was first suggested by a Wilson loop matrix in a strong
coupling expansion scheme (see figs. 4 and 5 of ref.\cite{matsuoka})
and appeared in above mentioned model \cite{P. Pennanen} of the
numerically evaluated Wilson loop matrix of the $SU(2)_c$
simulations. A full $SU(3)_c$ simulation \cite{green2} performed a
bit later again showed the need for $f$ model, though detailed
geometrical dependence of $f$ could not be studied in this 3 colour
lattice gauge study. But a later numerical lattice study \cite{V. G.
Bornyakov} by a reputed group of the {\em full} $2\times 2$ matrix
of the Wilson loops correlators and of "the interaction energy of
the confining strings in the static rectangular tetraquark system in
$SU(3)$ gluodynamics" was well modeled again by a surface model,
namely their {\em soap film} model that also incorporates (a flip
to) the multi $Y$ type linear potential emerging from recent
numerical simulations \cite{Hideo Suganuma}. The basis state overlap
$g$ \cite{V. G. Bornyakov} in the soap film model has a role similar
to the gluon field overlap factor $f$ of ref.\cite{P. Pennanen};
both $f$ and $g$ appear only in the off-diagonal terms of the
respective matrices ($N$ and $T$) of overlaps of the basis states.
Continuing on the spatial and space-time areas, it can be pointed
out that both kind of areas appear in eq.(13) of the ref.\cite{V. G.
Bornyakov} and are related to Wilson loops as earlier eq.(12) there
indicates.

Actually, above are models of the matrices of pure gluonic theory
Wilson loops. The diagonal terms in these matrices are time
evolutions of a tetraquark clustering and off-diagonal terms
\cite{matsuoka, V. G. Bornyakov} are for time evolution that start
from one tetraquark clustering (or topology) and end at another one.
The numerical evaluation of the off-diagonal Wilson loops has been
perhaps done in refs. \cite{P. Pennanen} (and previous ones by the
same group) and \cite{V. G. Bornyakov} only. But the diagonal Wilson
loops have been studied by many other groups, most familiar being
the studies reported in ref.\cite{Hideo Suganuma} and the previous
works by the same group. For one set of quark configurations, the
diagonal studies are limited to only one Wilson loop. In a sense,
this means limitation to only one state of the gluonic field as
well, namely the one with the least energy {\em to which} system
flips; if there are other states mentioned in the literature, these
are either for comparison purpose ({\em from which} the system
flips) or the excited state "contaminations". But a general study
should {\em actually} incorporate a variety of basis states and thus
include off-diagonal Wilson loops {\em as well}. The spatial area we
are working on appears only in the off-diagonal Wilson loops and
thus our work is not to be confused with the usual study of the
diagonal Wilson loops effects. It is to be admitted that works like
ref.\cite{Hideo Suganuma} have indicated improvements in both
evaluations and models of the diagonal Wilson loops and we have not
included these improvements in our model of the diagonal term. But
this is not a serious flaw, as a dynamical study \cite{J. Vijande}
using these improved diagonal models mentions in its conclusions and
comments that the "dynamics of (tetraquark) binding is dominated by
the simple flip-flop term", meaning that the essentially new
connected string (butterfly) term introduced through the work of
ref.\cite{Hideo Suganuma} is dynamically "not rewarding". It is to
be noted that our diagonal terms include both terms whose minimum is
the flip-flop term.

This advocates the $\exp(-k S)$ form of $f$ for static two quarks
and two antiquarks. For a comparison with actual (hadron)
experiments, we have to incorporate quark motion as well, possibly
through using quark wave functions. The resulting four-body
Schr\"{o}dinger equation can be solved, as in ref. \cite{J.
Weinstein},  variationally for the ground state of the system and
the effective meson-meson potentials. Alternatively, the Hamiltonian
emerging from the $q^2\bar{q}^2$ model has been diagonalized in the
simple harmonic oscillator basis \cite{B. Silvestre-Brac}, or was
sandwiched between the external meson wave functions to give a
transition amplitude of the {\em Born diagrams} \cite{T. Barnes,
E.S. Swanson} that is related to meson-meson phase shifts. We have
used a formalism (resonating group method \cite{wheeler}) that was,
for the $q^2\bar{q}^2$ system, originally \cite{B.Masud} used in a
way that allowed finding {\em inter-meson} dependence with the
dependence on a quark-antiquark distance being pre-fixed. The
formalism allows using the best available knowledge of a meson wave
function, though a simple Gaussian form for the wave functions and
correspondingly a quadratic quark-antiquark potential was used for
computational convenience. We have used this same formalism that can
be generalized to using realistic meson wave functions and finding
the inter-cluster dependence. But because of the additional
computational problems due to a totally non-separable exponential of
(a negative constant times) area in $f$, presently we had to also
pre-specify a plane-wave form of the inter-cluster dependence along
with still using Gaussian wave functions; effectively this means
using a Born approximation as well. We have pointed out, though,
that these wave function approximations are better than what their
naive impression conveys: The Gaussian dependence on the
quark-antiquark distance has the potential of resembling the
realistic meson wave functions through an adjustment of its
parameters \cite{ackleh}. And the plane wave form of the
inter-cluster dependence is justified through a feeble inter-cluster
(meson-meson) interaction noted in previous works \cite{J.
Weinstein, B.Masud, T. Barnes}; the meson-meson phase shifts
resulting from this work are also much less than a radian. {\em Only
by using the Born-approximation}, the resulting coupled integral
equations for inter-cluster wave functions could be decoupled in
this work. This decoupling allowed us to numerically calculate the
off-diagonal elements as nine-dimensional integrals for the
components of the eventual four position 3-vectors; only the overall
center-of-mass dependence could be analytically dealt with in a
trivial manner. Before this numerical integration, for the kinetic
energy terms we had to differentiate the area in $f$. The form of
area used in the detailed form of the $Q^2\bar{Q}^2$ model, that we
take from ref. \cite{P. Pennanen}, has square roots of the functions
of our position variables. Thus a differentiation of this area form
yields in denominators combinations of position variables that
become zero somewhere in the ranges of integrations to be later
done. The numerical evaluations of the resulting nine-dimensional
improper integrals is expected to be too demanding, as our initial
explorations indicated. Thus, for the to-be-differentiated right
$\sqrt{f}$ part of the some kinetic energy terms we replaced the
area by an approximated quadratic form whose differentiation does
not result in negative powers of the position variables.

We also find that the use of the $f$ factor in the new form  reduces the long
range meson-meson interaction and thus as usual solves the well known Van der
Walls force problem \cite{O. Morimatsu, J. Vijande} with the otherwise naive
sum of one gluon exchange pair-wise interaction. It has been said \cite{J.
Weinstein} that dynamically it is not a serious problem because of
the quark-antiquark pair creation and because of the wave function
damping of the large distance configurations. Though ref. \cite{green2} partly
incorporated both the quark-antiquark pair creation and the meson
wave functions and still showed a need for the $f$ factor, through
the present work we want to point out that the dynamical role of the
$f$ factor in meson-meson interactions is {\em not limited} to
solving Van der Walls force problem or pointing out \cite{masud}
otherwise over-binding in certain meson-meson systems. The $f$
factor points certain features (non-separability of $f$) of QCD that
are 1) indicated by lattice simulations and 2) can be compared with
actual experiments.

There have been recent hadron-level studies \cite{P. Bicudo}\cite{J.
Vijande} using the above mentioned improved models of the diagonal
Wilson loops. But the quark level limitations of the models
\cite{Fumiko Okiharu}\cite{Hideo Suganuma} mean similar limitations
for the hadron-level results: \cite{P. Bicudo}\cite{J. Vijande}
study the properties (like binding energy and {\em direct potential}
\cite{wong}) of the ground state itself (or in isolation), whereas
we aim to study the dynamical coupling of a tetraquark state to
other basis state(s) of the tetraquark system---essentially to the
other clustering or the exchanged channel. Thus, as we say in the
abstract, in addition to doing the phase-shift calculations for a
meson-meson scattering, we study a coupling that can affect even the
ground state itself through a second order perturbation theory
effect named polarization potential \cite{wong}. That is, after
including the quark mass differences, a meson-meson state may not be
degenerate with an exchanged channel and thus the coupling between
this state and the exchanged intermediate one (a hadron loop) may
help resolve the underlying structure of a possible meson-meson
state. Such a state may be a meson-meson molecule that can be formed
by the ground state. This also makes a study of the dynamical
coupling of a meson-meson channel to exchanged one worth pursuing.

In section 2 we have written the total state vector of the
$q^2\bar{q}^2$ system as in RGM, along with introducing the
Hamiltonian $H$ of the system without the $f$ factor and then
modifying $H$ through the $f$. In section 3 different position
dependent forms of $f$ have been described, including the
approximate forms that we had to use. In section 4 we have solved
the integral equations for a meson meson molecule in the absence of
spin degrees of freedom and with all equal quark masses. This section ends with a prescription
to find the phase shifts. In the last section we have presented the
numerical values of the phase shifts for different forms of $f$, for
different values of free parameter $k_{f}$ and for different values
of angle $\theta$ between $\textbf{P}_{1}$ and $\textbf{P}_{2}$.

\section{The $Q^2\bar{Q}^2$ Hamiltonian and the wave-function}
Using adiabatic approximation we can write the total state vector of
a system containing two quarks two antiquarks and the gluonic field
between them as a sum of the product of the quarks ($Q$ or $\bar Q$)
position dependent function
$\Psi_{g}(\textbf{r}_1,\textbf{r}_2,\textbf{r}_{\bar{3}},\textbf{r}_{\bar{4}})$
and the gluonic field state $|k\rangle_{g}$. $|k\rangle_{g}$ is
defined as a state which approaches $|k\rangle_{c}$  in the weak
coupling limit , with
$|1\rangle_{c}=|1_{1\bar{3}}1_{2\bar{4}}\rangle_{c}$,
$|2\rangle_{c}=|1_{1\bar{4}}1_{2\bar{3}}\rangle_{c}$ and
$|3\rangle_{c}=|\bar{3}_{12}3_{\bar{3}\bar{4}}\rangle_{c}$. In
lattice simulations of the corresponding (gluonic) Wilson loops it
is found that the lowest eigenvalue of the Wilson matrix, that is
energy of the lowest state, is always the same for both $2\times2$
and $3\times3$ matrices provided that $|1\rangle_{g}$ or
$|2\rangle_{g}$ has the lowest energy \cite{P. Pennanen}. The later
calculations \cite{V. G. Bornyakov} of the tetraquark system were
also done with a two level approximation. Taking advantage of these
observations, we have included in our expansion only two basis
states. As in resonating group method,
$\Psi_{g}(\textbf{r}_1,\textbf{r}_2,\textbf{r}_{\bar{3}},\textbf{r}_{\bar{4}})$
or
$\Psi_{g}(\textbf{R}_{c},\textbf{R}_{k},\textbf{y}_{k},\textbf{z}_{k})$
is written as product of known dependence on
$\textbf{R}_{c},\textbf{y}_{k},\textbf{z}_{k}$ and unknown
dependence on $\textbf{R}_{k}$. i.e.
$\Psi_{g}(\textbf{r}_1,\textbf{r}_2,\textbf{r}_{\bar{3}},\textbf{r}_{\bar{4}})=\Psi_{c}(\textbf{R}_{c})\chi_{k}(\textbf{R}_{k})\psi_{k}(\textbf{y}_{k},\textbf{z}_{k})$.
Here $\textbf{R}_{c}$ is the center of mass coordinate of the whole
system, $\textbf{R}_{1}$ is the vector joining the center of mass of
the clusters $(1,\overline{3})$ and $(2,\overline{4})$,
$\textbf{y}_{1}$ is the position vector of quark 1 with respect to
$\overline{3}$ within the cluster $(1,\overline{3})$ and
$\textbf{z}_{1}$ is the position vector of quark 2 with respect to
$\overline{4}$ within the cluster $(2,\overline{4})$. The same
applies to $\textbf{R}_{2}, \textbf{y}_{2}$ and $\textbf{z}_{2}$ for
the clusters $(1,\overline{4})$ and $(2,\overline{3})$. Similarly we
can define $\textbf{R}_{3}, \textbf{y}_{3}$ and $\textbf{z}_{3}$ for
the clusters $(1,2)$ and $(\overline{3},\overline{4})$. Or we can
write them in terms of position vector of the four particles (quarks
or antiquarks) as follow
\begin{equation}\textbf{R}_{1}=\frac{1}{2}(\textbf{r}_{1}+\textbf{r}_{\overline{3}}-\textbf{r}_{2}-\textbf{r}_{\overline{4}})\text{ , }\textbf{y}_{1}=\textbf{r}_{1}-\textbf{r}_{\overline{3}}\text{ and
}\textbf{z}_{1}=\textbf{r}_{2}-\textbf{r}_{\overline{4}},\label{e12}
\end{equation}
\begin{equation}\textbf{R}_{2}=\frac{1}{2}(\textbf{r}_{1}+\textbf{r}_{\overline{4}}-\textbf{r}_{2}-\textbf{r}_{\overline{3}})\text{ , }\textbf{y}_{2}=\textbf{r}_{1}-\textbf{r}_{\overline{4}}\text{ and
}\textbf{z}_{2}=\textbf{r}_{2}-\textbf{r}_{\overline{3}}\label{e13}
\end{equation}
and
\begin{equation}\textbf{R}_{3}=\frac{1}{2}(\textbf{r}_{1}+\textbf{r}_{2}-\textbf{r}_{\overline{3}}-\textbf{r}_{\overline{4}})\text{ , }\textbf{y}_{3}=\textbf{r}_{1}-\textbf{r}_{2}\text{ and
}\textbf{z}_{3}=\textbf{r}_{\overline{3}}-\textbf{r}_{\overline{4}}.\label{e14}
\end{equation}
Thus meson meson state vector in the restricted
gluonic basis is written as
\begin{equation}|\Psi(\textbf{r}_1,\textbf{r}_2,\textbf{r}_{\bar{3}},\textbf{r}_{\bar{4}};g)\rangle=
\sum_{k=1}^2|k\rangle_{g}\Psi_{c}(\textbf{R}_{c})\chi_{k}(\textbf{R}_{k})\xi_{k}(\textbf{y}_{k})\zeta_{k}(\textbf{z}_{k}).\label{e11}\end{equation}
Here $\xi_{k}(\textbf{y}_{k})=\frac{1}{(2\pi
d^{2})^{\frac{3}{4}}}\exp(\frac{-\textbf{y}_{k}^{2}}{4 d^{2}})$ and
$\zeta_{k}(\textbf{z}_{k})=\frac{1}{(2\pi
d^{2})^{\frac{3}{4}}}\exp(\frac{-\textbf{z}_{k}^{2}}{4 d^{2}})$.
These Gaussian forms of meson wave functions are, strictly speaking,
the wave functions of a quadratic confining potential. But, as
pointed out in text below fig. 1 of ref. \cite{ackleh}, the overlap
of a Gaussian wave function and the eigenfunction of the realistic
linear plus colombic potential can be made as close as 99.4\% by
properly adjusting its parameter $d$. A realistic value of $d$
mimicking a realistic meson wave function depends on the chosen
scattering mesons and thus is postponed to our future work
\cite{next}. Presently, to explore the qualitative implications of
the geometric features of the gluonic overlap factor $f$, we have
used in $\xi_{k}(\textbf{y}_{k})$ and $\zeta_{k}(\textbf{z}_{k})$ a
value $d=0.558$ fm defined by the relation
$d^{2}=\sqrt{3}R_{c}^{2}/2$ \cite{masud}, with $R_{c}=0.6$ fm being
the r.m.s. charge radius of the qqq system whose wave function is
derived by using the same quadratic confining potential.

As for the Hamiltonian, for $f$=1 the total Hamiltonian $H$ of our 4-particle system is
taken as \cite{John Weinstein}
\begin{equation}
\hat{H}=\sum_{i=1}^{\overline{4}}\Big[m_{i}+\frac{\hat{P}_{i}^{2}}{2m_{i}}\Big]+
\sum_{i<j}^{\overline{4}}
v(\textbf{r}_{ij})\mathbf{F}_{i}.\mathbf{F}_{j}.\label{e9}
\end{equation}
Our same constituent quark mass value $m=0.3\texttt{GeV}$ for all
quarks and antiquarks is one used in refs. \cite{B.Masud}, and our
kinetic energy operator is similarly non-relativistic; it is
included in our aims to compare with this work and isolate the
effects only due to a different expression for the $f$. In above
each of $\mathbf{F}_i$ has $8$ components $F_i^l=\lambda^l/2$,
$l=1,2,3,...,8$  and $F_i^{l*}=\lambda^{l*}/2$, $\lambda^{l}$ are
Gell-Mann matrices operating on the $i$-th particle; $l$ is shown as
a superscript only to avoid any possible confusion with subscript
$i$ which labels a particle.

For the pairwise $q\overline{q}$ potential, we have used a quadratic confinement
\begin{equation}v_{ij}=C
r_{ij}^{2}+\bar{C}
 \text{ with } i,j=1,2,\bar{3},\bar{4}.\label{e16}\end{equation} for
 exploratory study. While we have neglected short range coulomb like
 interactions as well as spin-dependent terms. Along with that non
 relativistic limit has also been taken. The model used by Vijande
 \cite{J. Vijande} is also restricted to these limits.
 As for the within-a-cluster dependence of the wave function, this use of
 the quadratic potential in place of the realistic Coulumbic plus linear may
 change the full wave function. In the within-a-cluster, this change of wave
 function is found to result in a change of an overlap integral from 100\% to 99.4\% only provided
 the parameter $d$ of the wave function is adjusted. Although the expression for $P_{c}$
 written immediately after eq.(\ref{e20}) suggests a way to connect the additional
 parameter of the full wave function with that of a cluster,
 it is difficult to make a similar overlap test for the full wave
 function. But there is no a priori reason to deny that at least the qualitative
 features (like a new kind of angle dependence mentioned in the results part)
 we want to point out using this quadratic confinement would survive in a more
 realistic calculation; a similar exploration of the $q^2\bar{q}^2$ system properties
 was first done \cite{J. Weinstein} using a quadratic confinement and later extended
 and improved calculation \cite{John Weinstein} with more realistic pair-wise interaction
 reinforced the $K\bar{K}$ results obtained initially through the quadratic confinement. It seems that
 a proper adjustment the parameters of the quadratic (or SHO) model can reasonably
 simulate a $q\bar{q}$ or even a $q^2\bar{q}^2$ system. In our case, this adjustment of
 the parameters can be done once a choice of actual scattering mesons is made in a
 formalism \cite{next} that incorporates spin and flavour degrees of freedom. But,
 as shown in fig. 2(b) of ref.\cite{J. Weinstein}, properties of the $q^2\bar{q}^2$ system
 may not be very sensitive to the actual values of the parameters and we expect our presently chosen
 values of the parameters to well indicate the essential features resulting from the
 non-separable form of the gluonic field overlap factor $f$.

For the central simple harmonic oscillator
potential of eq.(\ref{e16}), the above mentioned size $d$ in the eigenfunctions $\xi_{k}(\textbf{y}_{k})$
and $ \zeta_{k}(\textbf{z}_{k})$ is related to the quadratic
coefficient $C$ which thus is given a value of $-0.0097\texttt{GeV}^{3}$.

As in a resonating group calculation, we take only variations in the $\chi_{k}$ factor of the
total state vector of the system written in eq.(\ref{e11}). Setting the coefficients of
linearly independent arbitrary variations
$\delta\chi_{k}(\textbf{R}_{k})$ as zero and integrating out
$R_{c}$, $\langle\delta\psi\mid H-E_{c}\mid\psi\rangle=0$ from
eq.(\ref{e11}) gives

\begin{equation}
\sum_{l=1}^2\int d^3y_{k}d^3z_{k} \xi_{k}(\textbf{y}_{k})
\zeta_{{k}}(\textbf{z}_{k})_{g}\langle k\mid
H-E_{c}\mid\l\rangle_{g}
\chi_{l}(\textbf{R}_{l})\xi_{l}(\textbf{y}_{l})
\zeta_{l}(\textbf{z}_{l})=0, \label{e1}
\end{equation} for each of the $k$ values (1 and 2).
According to the (2 dimensional basis) model $I_{a}$ of ref.
\cite{P. Pennanen}, the normalization, potential energy and kinetic
energy matrices in the corresponding gluonic basis are
\begin{equation}N=\left(
            \begin{array}{cc}
              1 & \frac{1}{3} f \\
              \frac{1}{3} f  & 1 \\
            \end{array}
          \right),
\end{equation}

\begin{equation}V=\left(
            \begin{array}{cc}
              \frac{-4}{3} (v_{1\overline{3}}+v_{2\overline{4}}) & \frac{4}{9} f (v_{12}+v_{\overline{3}\overline{4}}-v_{1\overline{3}}-v_{2\overline{4}}-v_{1\overline{4}}-v_{2\overline{3}})\\
             \frac{4}{9} f (v_{12}+v_{\overline{3}\overline{4}}-v_{1\overline{3}}-v_{2\overline{4}}-v_{1\overline{4}}-v_{2\overline{3}}) & \frac{-4}{3} (v_{1\overline{4}}+v_{2\overline{3}}) \\
            \end{array}
          \right)
 \end{equation}

and

\begin{equation}
_g\langle k\mid K\mid\ l\rangle_{g}= N(f)_{k,l}^{\frac{1}{2}}
\Big(\sum_{i=1}^{\overline{4}}-\frac{\nabla_{i}^{2}}{2m}\Big)N(f)_{k,l}^{\frac{1}{2}}.\label{e15}
\end{equation}
This is the modification, through the $f$ factor, to the Hamiltonian
as much as we need it for the integral equations below in section 4
(that is only the modified matrix elements).

\section{Different forms of {\it f}}
Ref. \cite{P. Pennanen} supports through a comparison with numerical
lattice simulations a form of $f$ that was earlier \cite{O.
Morimatsu} suggested through a quark-string model extracted from the
strong coupling lattice Hamiltonian gauge theory. This is
 \begin{equation}f=\exp(-b_{s} k_{f} S)\label{e6},\end{equation}
S being the area of minimal surface bounded by external lines
joining the position of the two quarks and two antiquarks, and
$b_s=0.18\texttt{GeV}^{2}$ is the standard string tension
\cite{Isgur, Fumiko Okiharu}, $k_f$ is a dimensionless parameter
whose value of 0.57 was decided in ref. \cite{P. Pennanen} by a fit
of the simplest two-state area-based model (termed model Ia) to the
numerical results for a selection of $Q^2\bar{Q}^2$ geometries. It
is shown there \cite{P. Pennanen} that the parameters, including
$k_f$, extracted at this $SU(2)_{c}$ lattice simulation with
$\beta=2.4$ can be used directly in, for example, a resonating group
calculation of a four quark model as the continuum limit is achieved
for this value of $\beta$.

The simulations reported in ref. \cite{P. Pennanen} were done in the
2-colour approximation. But, for calculating the dynamical effects,
we use actual SU(3) colour matrix elements of ref. \cite{B.Masud}.
The only information we take from the computer simulations of ref.
\cite{P. Pennanen} is value of $k_{f}$. This describes a geometrical
property of the gluonic field (its spatial rate of decrease to zero)
and it may be the case that the geometrical properties of the
gluonic field are not much different for different number of
colours, as suggested for example by successes of the geometrical
flux tube model. Situation is more clear, though, for the mass
spectra and the string tension generated by the gluonic field: ref.
\cite{Teper} compare these quantities for $SU(2)_{c}$, $SU(3)_{c}$
and $SU(4)_{c}$ gauge theories in 2+1 dimensions and find that the
ratio of masses are, to a first approximation, are independent of
the number of colours. Their preliminary calculations in 3+1
dimensions indicate a similar trend. Directly for the parameter
$k_f$, appearing in the overlap factor $f$ studied in this work, the
similar conclusion can be drawn from a comparison of the mentioned
lattice calculations \cite{green2} on the interaction energy of the
two heavy-light $Q^2\bar{q}^2$ mesons in the realistic $SU(3)_{c}$
gauge theory with ref. \cite{P. Pennanen} that uses $SU(2)_{c}$. For
interpreting the results in terms of the potential for the
corresponding single heavy-light meson ($Q\bar{q}$), a Gaussian form
\begin{equation}f=\exp(-b_{s} k_{f}\sum_{i<j} r_{ij}^{2})\label{e7}\end{equation}
of the gluonic filed overlap factor $f$ is used in this ref.
\cite{green2} for numerical convenience and not the minimal area
form. But for a particular geometry, the two exponents (the minimal
area and the sum of squared distances) in these two forms of $f$ are
related and thus for a particular geometry a comparison of the
parameter $k_f$ multiplying area and corresponding (different!)
$k_f$ multiplying sum of squares in eq. (14) of ref. \cite{green2}
is possible. We note that, after correcting for a ratio of 8 between
the sum of distance squares (including two diagonals) and the area
for the square geometry, the colour-number-generated relative
difference for this geometry is just $5\%$: the coefficient is
$0.075\times 8=0.6$ multiplying sum of squared distances and $0.57$
multiplying the minimal area. But, as the precise form of $f$ is
still under development (the latest work \cite{V. G. Bornyakov} has
covered only a very limited selection of the positions of tetraquark
constituents) and the expression for the area in its exponent needs
improvement, it is not sure precisely what value of the $k_f$ best
simulates QCD and we have mainly worked with an approximate value of
0.5 that is also mentioned in ref.\cite{P. Pennanen} and is
numerically easier to deal with. (It is to be noted that the soap
film model of ref.\cite{V. G. Bornyakov} does not treat $k_f$ as a
variational parameter. If that is interpreted as fixing $k_f$ at 1,
this prescription might have been successful due to their selection
of quark configurations being limited to planar ones; a work
\cite{pennanen} by UKQCD that is limited to planar geometries also
favors a value closer to 1. But their more general work \cite{P.
Pennanen} resulted in a value of $k_f$ near 0.5.)

For the area as well, ref. \cite{P. Pennanen} used an approximation:
A good model of area of the minimal surface could be that given in
ref. \cite{green} as

\begin{equation}
S=\int_{0}^1 du\int_{0}^1 dv
|(u\textbf{r}_{1\overline{3}}+(1-u)\textbf{r}_{\overline{4}
2})\times(v\textbf{r}_{2\overline{3}}+(1-v)\textbf{r}_{\overline{4}
1})|
\end{equation}

(Work is in progress \cite{dawood}, to judge the surface used in
this model, and its area, from the point of differential geometry
and there are indications that this is quite close to a minimal
surface.) But the simulations reported in ref. \cite{P. Pennanen}
were carried out for the $S$ in eq.(\ref{e6}) being "the average
of the sum of the four triangular areas defined by the positions
of the four quarks". Although for the tetrahedral geometry the $S$
used in ref. \cite{P. Pennanen} is as much as 26 percent larger
than the corresponding minimal-like area of ref. \cite{green}, it
can be expected that their fitted value of $k_f$ is reduced to
partially compensate this over estimate of the $S$ area. Anyway,
as we are calculating the dynamical effects of the model of ref.
\cite{P. Pennanen}, we have used the form of $S$ that is used in
this work.

The area $S$ of ref. \cite{P. Pennanen} becomes (with a slight renaming) \\
\begin{eqnarray}S=\frac{1}{2}[{S(134)+S(234)+S(123)+S(124)}],\label{e22}\end{eqnarray} where
$S(ijk)$ is the area of the triangle joining the vertices of the
positions of the quarks labled as i,j and k. In the notation of
eqs.(\ref{e12}-\ref{e14}) this becomes
$S(134)=\frac{1}{2}|\textbf{y}_{1}\times
\textbf{z}_{3}|=\frac{1}{2}|(\textbf{R}_{2}+\textbf{R}_{3})\times(\textbf{R}_{1}-\textbf{R}_{2})|$,
$S(234)=\frac{1}{2}|\textbf{z}_{2}\times
\textbf{z}_{3}|=\frac{1}{2}|(\textbf{R}_{3}-\textbf{R}_{1})\times(\textbf{R}_{1}-\textbf{R}_{2})|$,
$S(123)=\frac{1}{2}|\textbf{y}_{3}\times
\textbf{z}_{2}|=\frac{1}{2}|(\textbf{R}_{1}+\textbf{R}_{2})\times(\textbf{R}_{3}-\textbf{R}_{1})|$
and $S(124)=\frac{1}{2}|\textbf{y}_{3}\times
\textbf{z}_{1}|=\frac{1}{2}|(\textbf{R}_{1}+\textbf{R}_{2})\times(\textbf{R}_{3}-\textbf{R}_{2})|$.
Written in terms of the rectangular components $(x_{1},y_{1},z_{1})$
of $\textbf{R}_{1}$, $(x_{2},y_{2},z_{2})$ of $\textbf{R}_{2}$ and
$(x_{3},y_{3},z_{3})$ of $\textbf{R}_{3}$, we have
\begin{eqnarray}
S(134)=\frac{1}{2}\Bigg\{\Big(x_2{}^2+y_2{}^2+z_2{}^2+2 \left(x_2 x_3+y_2 y_3+z_2 z_3\right)+x_3{}^2+y_3{}^2+z_3{}^2\Big)\hspace{3cm}\nonumber\\
\Big(x_1{}^2+y_1{}^2+z_1{}^2-2\left(x_1 x_2+y_1y_2+z_1 z_2\right)+x_2{}^2+y_2{}^2+z_2{}^2\Big)\hspace{3cm}\nonumber\\
-\Big(x_1 x_2+y_1y_2+z_1z_2-\left(x_2{}^2+y_2{}^2+z_2{}^2\right)+x_1
x_3+y_1 y_3+z_1z_3-\left(x_2
x_3+y_2y_3+z_2z_3\right)\Big)^{2}\Bigg\}^{\frac{1}{2}}. \label{e22}
\end{eqnarray}
Explicit rectangular expressions for $S(234), S(123)$ and $S(124)$ are similar.

This form of $S$ has square roots. For the $K.E$. part of the
Hamiltonian matrix (see eq.(\ref{e15})  ), we have to differentiate
an exponential of this square root. After differentiating we can
have negative powers of $S$ and when they will be integrated in the
latter stages can have singularities in the integrands resulting in
computationally too demanding nine-dimensional improper integrals.
Thus we have availed to ourselves an approximated $S$, named $S_a$,
which is a sum of different quadratic combinations of quarks
positions. We chose $S_{a}$ by minimizing $\int
d^{3}R_{1}d^{3}R_{2}d^{3}R_{3}(S-S_{a})^{2}$ with respect to the
coefficients of the quadratic position combinations; that is, these
coefficients are treated as variational parameters. The first
(successful) form  which we tried for $S_{a}$ was
  \begin{eqnarray}S_{a}=a \left(x_1{}^2+y_1{}^2+z_1{}^2\right)+b \left(x_2{}^2+y_2{}^2+z_2{}^2\right)
 +c \left(x_3{}^2+y_3{}^2+z_3{}^2\right)+{\acute d} x_1 x_2+\nonumber\\e y_1 y_2+f z_1 z_2+g x_2 x_3+h y_2 y_3+\acute{i} z_2 z_3+j x_1 x_3+k y_1 y_3+l z_1
 z_3.~~~~~~~~~~~~\label{e21}\end{eqnarray}
 This contained 12 variational parameters a,b,c,...,l.
Minimization gave values (reported with accuracy 4 though in the
computer program accuracy 16 was used)as
  \begin{eqnarray}
a = 0.4065, b = 0.4050, c = \ 0.3931, j = \ -0.0002, l = \ -0.0002.
\nonumber\end{eqnarray} In the reported accuracy other parameters
are zero. Here limits of integration were from -15 to 15 in
$\texttt{GeV}^{-1}$. We also tried $S_{a}$  as
$$\sum_{i=1}^3{a_{i}(x_{i}^{2}+y_{i}^{2}+z_{i}^{2})}+\sum_{i,j=1}^3{(b_{i,j}x_{i}y_{j}+c_{i,j}x_{i}z_{j}+d_{i,j}y_{i}z_{j})}+\sum_{i<j,j=2}^3{(e_{i,j}x_{i}x_{j}+f_{i,j}y_{i}y_{j}+g_{i,j}z_{i}z_{j})}$$
and
$$\sum_{i=1}^3{(l_{i}x_{i}^{2}+m_{i}y_{i}^{2}+n_{i}z_{i}^{2})}+\sum_{i,j=1}^3{(b_{i,j}x_{i}y_{j}+c_{i,j}x_{i}z_{j}+d_{i,j}y_{i}z_{j})}+\sum_{i<j,j=2}^3{(e_{i,j}x_{i}x_{j}+f_{i,j}y_{i}y_{j}+g_{i,j}z_{i}z_{j})},$$
with variational parameters being 39 and 45 respectively. Both the
latter forms gave the same result as we got with 12 variational
parameters, and hence this 12 parameter form was used in the section
below. This form gives dimensionless standard-deviation, defined as

$$\sqrt{\frac{\langle(S-S_{a})^{2}\rangle-(\langle
S-S_{a}\rangle)^{2}}{\langle S^{2}\rangle}},$$

 being approximately
equal to $21 \%$ . Here,

$$\langle
X\rangle=\frac{\int(X)d^{3}R_{1}d^{3}R_{2}d^{3}R_{3}}{\int(1)d^{3}R_{1}d^{3}R_{2}d^{3}R_{3}}.$$

As this is not too small, in our main calculations we have made a
minimal use of this further approximated area $S_a$ (only for the
to-be-differentiated right $\sqrt{f}$ part (see eq.\ref{e15} ) of
the kinetic energy term and here only for derivatives of the
exponent).
\section{Solving the integral equations}
 In eq.(\ref{e1}) for $k=l=1$ (a diagonal term), we used the linear independence
 of $\textbf{y}_{1}$, $\textbf{z}_{1}$ and $\textbf{R}_{1}$ (see
eq.(\ref{e12})) to take $\chi_{1}(\textbf{R}_{1})$ outside the
integrations w.r.t. $\textbf{y}_{1}$ and $\textbf{z}_{1}$. For the
off-diagonal term with $k=1$ and $l=2$ we replaced $\textbf{y}_{1}$
and $\textbf{z}_{1}$ with $\textbf{R}_{2}$ and $\textbf{R}_{3}$,
with Jacobian of transformation as 8. For regulating the space
derivatives of the exponent of $f$ (see the three sentences
immediately following eq.(\ref{e22}) above) we temporarily replaced
$S$ in it by its quadratic approximation $S_{a}$. As a result, we
obtained the following equation:
\begin{eqnarray}
\bigg(\frac{3\omega}{2}-\frac{1}{2\mu_{12}}\nabla^{2}_{R_{1}}+24
C_{o} d^{2}-\frac{8}{3}
\overline{C}-E_{c}+4m\bigg)\chi_{1}(\textbf{R}_{1})~~~~~~~~~~~~~~~~~~~~~~~~~~~~~~~~~~~~~~~~~~~~~~~~~~~~~~~~~~~~~~~~~~\nonumber\\+
 \int
d^3R_{2}d^3R_{3}\exp\Big(-b_{s}k_{f}S\Big)
\exp\Bigg(-\frac{R^{2}_{1}+R^{2}_{2}+2R^{2}_{3}}{2d^{2}}\Bigg)
\Bigg[-\frac{8}{6m(2\pi
d^{2})^{3}}g_{1}\exp\Big(\frac{1}{2}b_{s}k_{f}S\Big)~~~~~~~~~~~~~~~~~~~~\nonumber\\
\exp\Big(-\frac{1}{2}b_{s}k_{f}S_{a}\Big) +\frac{32}{9(2\pi
d^{2})^{3}}\Big(-4CR^{2}_{3}-2\overline{C}\Big)-
\frac{8(E_{c}-4m)}{3(2\pi
d^{2})^{3}}\Bigg]\chi_{2}(\textbf{R}_{2})=0,~~~~~~~~~~~~~\label{e2}
\end{eqnarray}
with, written up to accuracy 4,
 \begin{eqnarray}
g_{1}=-1.4417+0.0258 x_1{}^2+0.0258 x_2{}^2\
+0.0254 x_3{}^2+0.0258 y_1{}^2+\nonumber\\
 0.0258 y_2{}^2 +0.0254 y_3{}^2+\
 0.0258 z_1{}^2+ 0.0258 z_2{}^2+
 0.0254 z_3{}^2.\label{e4}
 \end{eqnarray}
For the consistency of $\xi_{k}(\textbf{y}_{k})$ and
$\zeta_{k}(\textbf{z}_{k})$ with eq.(\ref{e16})
$\omega=1/md^{2}=0.416\texttt{GeV}$. For convenience in notation we
take $C_{o}=-C/3$. Here in the first channel for $k=1$ the
constituent quark masses has been replaced by the reduced mass
$\mu_{12}=M_{1}M_{2}/(M_{1}+M_{2})$, where $M_{1}$ and $M_{2}$ are
masses of hypothetical mesons; a similar replacement has been done
in ref.\cite{B.Masud}.

At this stage we can fit $\bar C$ to a kind of "hadron spectroscopy"
for our equal quark mass case:

For the large separation there is no interaction between $M_{1}$ and
$M_{2}$. So the total center of mass energy in the large separation
limit will be the sum of kinetic energy of relative motion and
masses of $M_{1}$ and $M_{2}$ i.e. in the limit
$R_{1}\longrightarrow\infty$ we have
\begin{equation}
\Bigg[-\frac{1}{2\mu_{12}}\nabla^{2}_{R_{1}}+M_{1}+M_{2}\Bigg]\chi_{1}(\textbf{R}_{1})=E_{c}\chi_{1}(\textbf{R}_{1}).\label{e8}
\end{equation}
By comparing, in this limit, eq.(\ref{e8}) and eq.(\ref{e2}) we have
$M_{1}+M_{2}=4m+3\omega-8\bar{C}/3$. (A use of the first term of
eq.(\ref{e16}) for the colour-basis diagonal matrix element of
eq.(\ref{e9}) gives $-4C/3=\mu \omega^2/2=\mu\omega/2m d^{2}$,
giving $24
 C_{o}d^{2}=3\omega/2$ for the reduced mass $\mu$ of a pair of equal mass quarks being $m/2$  ; the
diagonal elements in any form of the $f$ model for the gluonic basis
are the same as those for the colour basis.) By choosing
$M_{1}+M_{2}=3\omega$ we have $\bar{C}=3m/2=0.45\texttt{GeV}$. This
choice of the hypothetical meson masses is the one frequently used
in ref. \cite{B.Masud} for an illustration of the formalism; when we
incorporate flavour and spin dependence \cite{next} the same fit,
something like in ref. \cite{masud}, we plan to fit our quark masses
to actual meson spectroscopy. We can then choose to fit even the
parameter $C$ or $C_0$ of our potential model to hadron spectroscopy
rather than deciding it, as in ref. \cite{B.Masud} and the present
work, through a combination of baryon radii and harmonic oscillator
model. But we do not see any reason why the qualitative effects (for
example, an angle dependance, see the section below) pointed out
through the present work should disappear for a phenomenologically
explicit case.

Completing our integral equations before finding a solution for two
$\chi 's$, for $k=2$ in eq.(\ref{e1})  we took
$\chi_{2}(\textbf{R}_{2})$ outside of integration for the diagonal
term, for the off-diagonal term we replaced $\textbf{y}_{2}$ and
$\textbf{z}_{2}$ by $\textbf{R}_{1}$ and $\textbf{R}_{3}$ and
replaced $S$ by $S_a$. This resulted in
\begin{eqnarray}
\bigg(\frac{3\omega}{2}-\frac{1}{2\mu_{34}}\nabla^{2}_{R_{2}}+24
C_{o} d^{2}-\frac{8}{3}
\overline{C}-E_{c}+4m\bigg)\chi_{2}(\textbf{R}_{2})~~~~~~~~~~~~~~~~~~~~~~~~~~~~~~~~~~~~~~~~~~~~~~~~~~~~~~~~~~~~~~~~~~~~~~\nonumber\\+
\int d^3R_{1}d^3R_{3}
\exp\Big(-b_{s}k_{f}S\Big)\exp\Bigg(-\frac{R^{2}_{1}+R^{2}_{2}+2R^{2}_{3}}{2d^{2}}\Bigg)
\Bigg[-\frac{8}{6m(2\pi d^{2})^{3}} g_{1}
\exp\Big(\frac{1}{2}b_{s}k_{f}S\Big)~~~~~~~~~~~~~~~~~~~~~~~\nonumber\\
\exp\Big(-\frac{1}{2}b_{s}k_{f}S_{a}\Big)+\frac{32}{9(2\pi
d^{2})^{3}}\Big(-4CR^{2}_{3}-2\overline{C}\Big)-
\frac{8(E_{c}-4m)}{3(2\pi
d^{2})^{3}}\Bigg]\chi_{1}(\textbf{R}_{1})=0.~~~~~~~~~~~~~~~~~~~~~\label{e3}
\end{eqnarray}

In the 2nd channel, for $k=2$, the constituent quark masses are
replaced by the reduced mass $\mu_{34}=M_{3}M_{4}/(M_{3}+M_{4})$,
where $M_{3}$ and $M_{4}$ are masses of hypothetical mesons.

 Now we solve our two integral
equations. As our space derivatives have been regularized, we no
longer need further-approximated $S_a$ and we replace this by the
original $S$ in eq.(\ref{e3}). Below we take Fourier transform of
eq.(\ref{e2}). This gives us a nine dimensional integral of, amongst
others, $\exp(-b_{s}k_{f}S)$. Non-separability of $S$ did not allow
us to formally solve the two integral equations for a non-trivial
solution for $\chi_1$ and $\chi_2$ as in ref. \cite{B.Masud} and we
had to pre-specify a form for $\chi_{2}(\textbf{R}_{2})$ in
eq.(\ref{e2}) and of $\chi_{1}(\textbf{R}_{1})$ in eq.(\ref{e3}).
(As long as all the functions, including the meson wave functions
and the gluonic field overlap factor $f$, are separate in
$\textbf{R}_{1}$ and $\textbf{R}_{2}$ , we can everywhere replace
$\chi_1$ and $\chi_2$ by their analytical integrals which themselves
simply multiply if $\chi_1$ and $\chi_2$ do, can solve the resulting
linear equations for these integrals and can write the $T$ matrices
and phase shifts directly in terms of these integrals. This is what
is done in ref.\cite{B.Masud}\cite{masud}, but it is hard to think
how to generalize this very specialized technique to a case like us
where the $f$ factor is not separable in $\textbf{R}_{1}$ and
$\textbf{R}_{2}$.) Compelled to use, thus, Born approximation
(something already in use \cite{T. Barnes} for meson-meson
scattering; our numerical results mentioned below also justify its
use here) for this we used the solutions of eqs.(\ref{e2}) and
(\ref{e3}) in absence of interactions (say by letting $k_f$ approach
to infinity, meaning $f=0$) for $\chi_{1}(\textbf{R}_{1})$ and
$\chi_{2}(\textbf{R}_{2})$. We chose the coefficient of these plane
wave solutions so as to make $\chi_{1}(\textbf{R}_{1})$ as Fourier
transform of $\delta({P}_{1}-{P}_{c}(1))/P_{c}^{2}(1)$ and
$\chi_{2}(\textbf{R}_{2})$ as Fourier transform of
$\delta({P}_{2}-{P}_{c}(2))/P_{c}^{2}(2)$, with $P_{c}(1)$ and
$P_{c}(2)$ defined below just after eq.(\ref{e20}). Thus we used
\begin{equation}\chi_{2}(\textbf{R}_{2})=\sqrt{\frac{2}{\pi}}\exp\Big(i
\textbf{P}_{2}.\textbf{R}_{2}\Big)\label{e18}\end{equation} inside
the integral to get one equation (after a Fourier transform with
respect to $R_{1}$ and kernel $e^{i \textbf{P}_{1}.\textbf{R}_{1}}$)
as
\begin{eqnarray}
\Big(3\omega+\frac{P_{1}^{2}}{2\mu_{12}}-E_{C}\Big)\chi_{1}(\textbf{P}_{1})=~~~~~~~~~~~~~~~~~~~~~~~~~~~~~~~~~~~~~~~~~~~~~~~~~~~~~~~~~~~~~~~~~~~~~~~\nonumber\\-\sqrt{\frac{2}{\pi}}\frac{1}{(2\pi)^{\frac{3}{2}}}\int
d^3R_{1}d^3R_{2}d^3R_{3}
\exp\Bigg\{i\Big(\textbf{P}_{1}.\textbf{R}_{1}+\textbf{P}_{2}.\textbf{R}_{2}\Big)\Bigg\}
\nonumber\\\exp\Big(-b_{s}k_{f}S\Big)
\exp\Big(-\frac{R^{2}_{1}+R^{2}_{2}+2R^{2}_{3}}{2d^{2}}\Big)
\nonumber \\ \Bigg[-\frac{8}{6m(2\pi
d^{2})^{3}}g_{1}+\frac{32}{9(2\pi
d^{2})^{3}}\Big(-4CR^{2}_{3}-2\overline{C}\Big)-
\frac{8(E_{C}-4m)}{3(2\pi d^{2})^{3}}\Bigg],\label{e10}
\end{eqnarray}
with $\chi_{1}(\textbf{P}_{1})$ being Fourier transform of
$\chi_{1}(\textbf{R}_{1})$. The formal solution~\cite{B.Masud} of
eq.(\ref{e10}) can be written as
\begin{eqnarray}
\chi_{1}(\textbf{P}_{1})=\frac{\delta({P}_{1}-{P}_{c}(1))}{P_{c}^{2}(1)}-\frac{1}{\Delta_{1}(P_{1})}\frac{1}{16\pi^{5}d^{6}}\int
d^3R_{1}d^3R_{2}d^3R_{3}\exp\Bigg\{i\Big(\textbf{P}_{1}.\textbf{R}_{1}+\textbf{P}_{2}.\textbf{R}_{2}\Big)\Bigg\}
~~~~~~~~~~~~~~~~~~~\nonumber\\
\exp\Big(-b_{s}k_{f}S\Big)\exp\Bigg(-\frac{R^{2}_{1}+R^{2}_{2}+2R^{2}_{3}}{2d^{2}}\Bigg)
\Bigg[-\frac{8}{6m}g_{1}+\frac{32}{9}\Big(-4CR^{2}_{3}-2\overline{C}\Big)-
\frac{8}{3}(E_{C}-4m)\Bigg],
\end{eqnarray}
with
$$\Delta_{1}(P_{1})=\frac{P_{1}^{2}}{2\mu_{12}}+3\omega-E_{c}-i\varepsilon.$$

 If we choose \emph{x}-axis along $\textbf{P}_{1}$ and choose
\emph{z}-axis in such a way that \emph{xz}-plane becomes the plane
containing $\textbf{P}_{1}$ and $\textbf{P}_{2}$, the above equation
becomes
\begin{eqnarray}
\chi_{1}(\textbf{P}_{1})=\frac{\delta({P}_{1}-{P}_{c}(1))}{P_{c}^{2}(1)}-\frac{1}{\Delta_{1}(P_{1})}F_{1},
\label{e19}
\end{eqnarray}
where, in the notation of eq.(\ref{e22}),

\begin{eqnarray}
F_{1}=\frac{1}{16\pi^{5}d^{6}}\int_{-\infty}^{\infty}dx_{1}dx_{2}dx_{3}dy_{1}dy_{2}dy_{3}dz_{1}dz_{2}dz_{3}\exp\Big\{i
P(x_{1}+x_{2}\cos\theta+z_{2}\sin\theta)\Big\}
\exp\Big(-b_{s}k_{f}S\Big)~~~~~~~~~~~~~~\nonumber\\
\exp\Bigg\{-\frac{x^{2}_{1}+y^{2}_{1}+z^{2}_{1}+x^{2}_{2}+y^{2}_{2}+z^{2}_{2}+2(x^{2}_{3}+y^{2}_{3}+z^{2}_{3})}{2d^{2}}\Bigg\}\nonumber\\
\left[-\frac{8}{6m}g_{1}+\frac{32}{9}\Big\{-4C(x^{2}_{3}+y^{2}_{3}+z^{2}_{3})-2\overline{C}\Big\}-\frac{8}{3}(E_C-4m)\right].\label{e40}
\end{eqnarray}
Here $\theta$ is the angle between $\textbf{P}_{2}$ and
$\textbf{P}_{1}$ and because of elastic scattering $P_{1}=P_{2}=P$.
From eq.(\ref{e19}) we can write, as in ref. \cite{B.Masud}, the
$1,2$ element of the T-matrix as
\\ \begin{eqnarray}
T_{12}=2\mu_{12} \frac{\pi}{2}P_{c} F_{1}.\label{e20}
\end{eqnarray}  Here $P_{c}=P_{c}(2)=P_{c}(1)=\sqrt{2\mu_{12} (E_{c}-(M_{1}+M_{2}))}$ and $M_{1}=M_{2}=3\omega/2$; see paragraph after eq.(\ref{e8}).
Using the relation $$s=I-2i T=\exp(2i\Delta)$$  or $$\left(
        \begin{array}{cc}
          1 & 0 \\
          0 & 1 \\
        \end{array}
      \right)
-2i \left(
          \begin{array}{cc}
            T_{11} & T_{12} \\
            T_{21} & T_{22} \\
          \end{array}
        \right)=\left(
        \begin{array}{cc}
          1 & 0 \\
          0 & 1 \\
        \end{array}
      \right)
+2i \left(
          \begin{array}{cc}
           \delta_{11} & \delta_{12} \\
           \delta_{21} & \delta_{22} \\
          \end{array}
        \right)
$$  between $s$ matrix and the $T$ matrix (actually in the form of
elements $\delta_{ij}=-T_{ij}$ for $i,j=1,2$) we got different
results for phase shifts for different values of center of mass
kinetic energy $T_c$ and the angle $\theta$ between $\textbf{P}_{1}$
and $\textbf{P}_{2}$; we have used the Born approximation to neglect
higher powers in the exponential series. We also probed different
values of the parameter $k_f$.

For a comparison, we also did the much less time consuming (but
approximate) calculation using $S_{a}$ in place of $S$ in
eq.(\ref{e10}). This allowed us separating the $9$ variables
dependence of the integrand as a product, resulting in three triple
integrals to be only multiplied, making the convergence very fast in
the numerical computation of the integral. Thus we had instead

\begin{eqnarray}
\chi_{1}(\textbf{P}_{1})=\frac{\delta({P}_{1}-{P}_{c}(1))}{P_{c}^{2}(1)}-\frac{1}{\Delta_{1}(P_{1})}{\emph{F}},
\hspace{1in}
\label{e5} \text{with} \end{eqnarray}  \\
\begin{eqnarray}
\emph{F}=\frac{1}{16\pi^{5}d^{6}}\Bigg[\int_{-\infty}^{\infty}
dx_{1}dx_{2}dx_{3}\Bigg\{f_{1}(x_{1},x_{2},x_{3})
\exp\Bigg[-\frac{{x_{1}}^2+x_{2}^{2}+2x_{3}^{2}}{2d^{2}}~~~~~~~~~~~~~~~~~~~~~~~~~~~~~~~~~~~~~~~~~~\nonumber\\-b_{s}k_{f}
\Big(a x_{1}^{2}+dx_{1}x_{2}+jx_{1}x_{3}+b
x_{2}^{2}+gx_{2}x_{3}+cx_{3}^{2}\Big)+i
P(x_{1}+x_{2}\cos\theta)\Bigg]\Bigg\}Q(y)\times Q(z) + \nonumber\\
\int_{-\infty}^{\infty}
dy_{1}dy_{2}dy_{3}\Bigg\{f_{2}(y_{1},y_{2},y_{3})
\exp\Bigg[-\frac{{y_{1}}^2+y_{2}^{2}+2y_{3}^{2}}{2d^{2}}~~~~~~~~~~~~~~~~~~~~~~~~~~~~~~~~~~~~~~~~~~\nonumber\\-b_{s}k_{f}
\Big(a y_{1}^{2}+ey_{1}y_{2}+ky_{1}y_{3}+b
y_{2}^{2}+hy_{2}y_{3}+cy_{3}^{2}\Big)\Bigg]\Bigg\}Q(x)\times
Q(z)~~~\nonumber\\ + \int_{-\infty}^{\infty}
dz_{1}dz_{2}dz_{3}\Bigg\{f_{3}(z_{1},z_{2},z_{3})
\exp\Bigg[-\frac{{z_{1}}^2+z_{2}^{2}+2z_{3}^{2}}{2d^{2}}~~~~~~~~~~~~~~~~~~~~~~~~~~~~~~~~~~~~~~~~~~~~\nonumber\\-b_{s}k_{f}
\Big(a z_{1}^{2}+fz_{1}z_{2}+lz_{1}z_{3}+b
z_{2}^{2}+\acute{i}z_{2}z_{3}+cz_{3}^{2}\Big)+i
Pz_{2}\sin\theta\Bigg]\Bigg\}Q(x)\times Q(y) \Bigg].
\end{eqnarray}

Here

$f_{1}(x_{1},x_{2},x_{3})=-\frac{8}{6m} \Big(-1.4417+0.0258
{x_{1}}^2+0.0254 {x_{2}}^2-{4.1914\times 10^{-7}} x_{1} x_{3}+0.0258
{x_{3}}^2\Big)+ \\~~~~~~~~~~~~~~~~~~~~~~~~~~~ \frac{32}{9} \Big(-4 C
x_{3}^{2}-2 \overline{C}\Big)-\frac{8}{3} \Big(E_{c}-4 m\Big)$,

$f_{2}(y_{1},y_{2},y_{3})=-\frac{8}{6m}\Big(0.0258 {y_{1}}^2+0.0254
{y_{2}}^2-{4.1914\times 10^{-7}} y_{1} y_{3}+ 0.0258 {y_{3}}^2
\Big)-\frac{128}{9} C y_{3}^{2}$,

$f_{3}(z_{1},z_{2},z_{3})=-\frac{8}{6m}\Big(0.0258 {z_{1}}^2+0.0254
{z_{2}}^2+{5.1396\times 10^{-6}} z_{1} z_{3}+0.0258
{z_{3}}^2\Big)-\frac{128}{9} C z_{3}^{2}$,\\

$Q(x)=\int_{-\infty}^{\infty}
dx_{1}dx_{2}dx_{3}\exp\Big[-\frac{{x_{1}}^2+x_{2}^{2}+2x_{3}^{2}}{2d^{2}}-b_{s}k_{f}
\Big(a x_{1}^{2}+dx_{1}x_{2}+jx_{1}x_{3}+b
x_{2}^{2}+gx_{2}x_{3}+cx_{3}^{2}\Big)+\\
~~~~~~~~~~~~~~~~~~~~~~~~~~~~~~~~~~~~~i
P(x_{1}+x_{2}\cos\theta)\Big]$,

$Q(y)=\int_{-\infty}^{\infty}
dy_{1}dy_{2}dy_{3}\exp\Big[-\frac{{y_{1}}^2+y_{2}^{2}+2y_{3}^{2}}{2d^{2}}-b_{s}k_{f}
\Big(a y_{1}^{2}+ey_{1}y_{2}+ky_{1}y_{3}+b
y_{2}^{2}+hy_{2}y_{3}+cy_{3}^{2}\Big)\Big]$\\

$Q(z)=\int_{-\infty}^{\infty}
dz_{1}dz_{2}dz_{3}\exp\Big[-\frac{{z_{1}}^2+z_{2}^{2}+2z_{3}^{2}}{2d^{2}}-b_{s}k_{f}
\Big(a z_{1}^{2}+fz_{1}z_{2}+lz_{1}z_{3}+b
z_{2}^{2}+\acute{i}z_{2}z_{3}+cz_{3}^{2}\Big)+i
Pz_{2}\sin\theta\Big].$\newline

For this choice of $S$, we also calculated the phase shifts that are
reported in the next section.

By treating eq.(\ref{e3}) in the same fashion as that of
eq.(\ref{e2}) and using the Born approximation
\begin{equation}\chi_{1}(\textbf{R}_{1})=\sqrt{\frac{2}{\pi}}\exp\Big(i
\textbf{P}_{1}.\textbf{R}_{1}\Big)\label{e17}\end{equation} it was
checked that the results for phase shifts remain same. Actually
eq.(\ref{e3}) and eq.(\ref{e2}) become identical if we interchange
$\textbf{R}_{1}$ and $\textbf{R}_{2}$.

\section{Results and conclusion}

\begin{figure}
\begin{center}
\epsfig{file=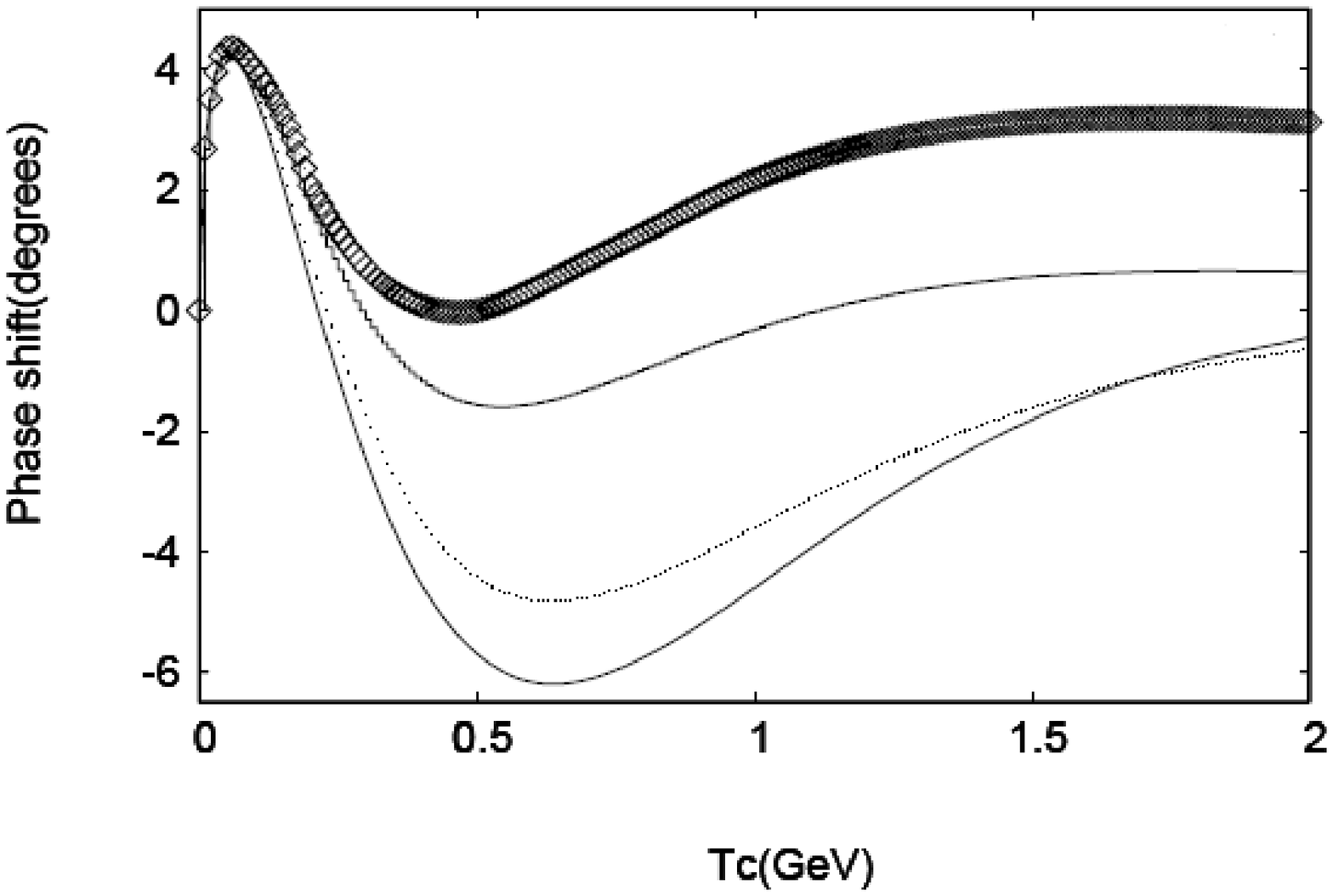,width=0.75\linewidth,clip=} \caption{For
$k_{f}=0.5$, the comparison of phase shifts for different values of
$\theta$ using $S$ in $f$.  The graph with points is for $\theta=0$,
with steps for $\theta=\frac{\pi}{6}$ and with dots only is for
    $\theta=\frac{\pi}{3}$. The graph with lines is for
    $\theta=\frac{\pi}{2}$; here the data is equally spaced as in the other graphs but data points are joined.} \label{graph1}
\end{center}
\end{figure}

Fig.\ref{graph1} shows our results, with $k_f$  defined by
eq.(\ref{e6}) taken as 0.5, for the phase shifts for a selection of
center of mass kinetic energies for different angles between
$\textbf{P}_{1}$ and $\textbf{P}_{2}$ (Some numerical uncertainty
appears at $0.15\texttt{GeV}$ for $\theta=0$. When we further
explored the region between $0.14\texttt{GeV}$ and
$0.16\texttt{GeV}$ there appeared fluctuations in the results. For
smoothness in graph we have neglected the data point at
$0.15\texttt{GeV}$ in fig.\ref{graph1} at $\theta=0$ and
interpolated data points are taken there.)

We found no numerical fluctuations for kinetic energies above
$0.16\texttt{GeV}$, and thus we conclude that in this kinematical
range the scattering angle has large effect on phase shifts,
indicating a true gluonic field effect. (The origin of this angle dependence is the exponent
  $S$ which is essentially in a model of $W_{12}$
Wilson loop, a pure gluonic theory expectation value; we do not get
any angle dependence if this $S$ is not used. So the angle
dependence emerges from gluonic field related to the area law, and
is thus a QCD effect.) By increasing the scattering angle the phase
shifts become large. We noted that a faster convergence of the
nine-dimensional integration (see eq.(\ref{e40})) for large kinetic
energy values was possible for smaller values of the parameter
$k_f$; for a decrease in $k_{f}$ of 0.1 the CPU time reduced at
least three times to that for the previous value. Thus we used the
smaller value of $k_f=0.5$ mentioned in ref.\cite{P. Pennanen} to
get phase shifts for a larger set of kinetic energies resulting in
smoother graphs. For the above mentioned value
$C=-0.0249281\texttt{GeV}^{3}$ (meaning
$\omega=0.665707\texttt{GeV}$ and $d=0.441357$ fm) used in
ref.\cite{ackleh} (giving the 99.4\% overlap of the wave functions)
we found that, at $T_{c}=0.1\texttt{GeV}$, there is about a $1$
degree change in phase shift for a $30$ degree change in scattering
angle $\theta$, even larger in magnitude than the phase shifts of
fig.\ref{graph1} for the corresponding $T_c$ given by our routine
value $d=0.556$ fm. So we can say that the characteristic feature of
angle dependence will remain if we study the scattering of some
realistic meson meson system by taking sizes of mesons accordingly
and adjust the parameters of the Gaussian wave functions to simulate
realistic linear plus Columbic potential eigenfunctions.

 For a comparison with the other crude forms of $f$ previously used, we show in the following fig.\ref{graph2} the
average of these $k_f=0.5$ phase shifts over our selection of angle
$\theta$ values along with the corresponding phase shifts for other
forms of $f$ i.e. exponent in $f$ being proportional to $S_{a}$,
$\sum_{i<j} r_{ij}^{2}$ and zero; the phase shifts were found to be
independent of the angle $\theta$ for all these older forms of $f$
and hence there was no need to take any angle-average for these
other forms. This figure shows that in comparison to $k_{f}=0$ (sum
of two body potential model) we get relatively very small coupling
with $S$, $S_{a}$ and Gaussian form in $f$. The introduction of a
many-body interaction in the previous (Gaussian) form of $f$
resulted in a reduced meson-meson interaction. In ref.
\cite{B.Masud} this reduction was noted as decreased meson-meson
phase shifts. So there are less chances of making a bound state with
modifications in sum of two body approach i.e. the inclusion of
gluonic field effects significantly decrease coupling between two
mesons in a $q^{2}\overline{q}^{2}$ system. Phase shifts are much
less than $1$ radian which indicates the validity of Born
approximation. The phase shifts we get are lesser than reported by
others who have used Born approximation \cite{T. Barnes} but not
used $f$ factor in off-diagonal terms.

It is to be noted that the $S_a$ form also does not result in any
angle dependence, although in contrast to the $f=1$ and Gaussian
form there is apparently no a priori reason to expect such an angle
independence for the use of $S_a$. This may be because $S_a$ is
almost Gaussian with a little mixture of $x_1 x_3$ and $z_1 z_3$
terms (see eq.(\ref{e21}) and the parameter values reported just
below it) or because $S_a$ can be converted to a Gaussian form by a
completing of squares. As for a comparison of the $S_a$ phase shifts
with the angle-average of the $S$ phase shifts, it can be pointed
out that the height of phase shift with $S_a$ became less than that
with original $S$ but the shape remains identical. Perhaps this
indicates that $S_{a}$ simulates well some variations resulting from
the original $S$ form. In fig.\ref{graph2}, if we compare graph of
Gaussian form with the angle averaged phase shifts using $S$ in $f$
we find that as compared to Gaussian form the graph of other forms
is closer to $k_{f}=0$, though the height of graph with $k_{f}=0$ is
still very large as compared to both Gaussian form of $f$ and that
of  $S$ in $f$; see fig.\ref{graph7} which clarifies any ambiguity,
if so, in fig.\ref{graph2} about the $k_f$=0 results.

Fig.\ref{graph3} reports most of the results for the higher values
0.57 of $k_f$ mentioned in ref. \cite{P. Pennanen}. This value is
more precise for their form of model and the crude area expression
in the exponent in it. But our numerical calculations for this value
turned out to be more demanding and thus were done for a smaller
selection of kinetic energies. The numerical uncertainties for this
value are for $\theta=\frac{\pi}{2}$ and the kinetic energy between
$0.11\texttt{GeV}$ and $0.12\texttt{GeV}$; the results for this
value of $\theta$ are in fig.\ref{graph5}.

A value of $k_f=1.0$ higher than 0.5 and 0.57 mentioned above has
been reported in ref. \cite{pennanen}. Although this work analyzes a
relatively limited collection of geometries (only squares and tilted
rectangles), we have tried to see effects of using a higher $k_f$.
The numerical problems for large $k_f$ implied in the above
mentioned numerical convenience for smaller $k_f$ did not allow us
to get results for $k_f$=1 in a manageable time  even for
$T_{c}=0.1\texttt{GeV}$. The best we could do was to do a number of
calculations for $k_f=0.8$; the resulting phase shifts from these
are shown in fig.\ref{graph4} except for the phase shifts for
$\theta=\pi/2$ that are reported in fig.\ref{graph6}, showing some
numerical uncertainties at the kinetic energy=$0.13\texttt{GeV}$ and
at $0.49-0.51\texttt{GeV}$. Based on fig.\ref{graph4} and
fig.\ref{graph6}, we expect that for higher values of $k_{f}$ the
results will remain qualitatively same and do not expect any new
feature to emerge for $k_{f}=1$.

We did a partial wave analysis for the phase shifts in fig.
\ref{graph1}. For this we have projected our angle dependence on
$m=0$ spherical harmonics. The angle dependence is independent of
azimuthal angle $\phi$ so the partial wave expansion will only
contain terms independent of angle $\phi$ or terms with m=0. This
analysis shows that below $0.2\texttt{GeV}$ the S-wave is very much
dominant as can be seen by a sharp rise of graphs towards lower side
in fig. \ref{graph8} and fig. \ref{graph9} near $0.2\texttt{GeV}$
and it is also justified from fig. \ref{graph1} which shows a
$\theta$ independence below $0.2\texttt{GeV}$ i.e. purely an S-wave.
Our partial wave analysis shows that only even partial waves are
present. Furthermore partial wave phase shifts decrease as we go
from S-wave to I-wave, as is clear from large reciprocals in fig.
\ref{graph9}, and they become negligible as compared to
corresponding S-wave phase shifts as we go beyond G-wave. So here in
figs. (\ref{graph8},\ref{graph9}) only S/D and S/G ratios are
plotted. The reason for the absence of odd partial waves is that our
phase shifts are symmetric around $\theta=\frac{\pi}{2}$ and the
product of an even and odd function is an odd function giving rise
to a zero result after integration. Thus phase shift is different
for different angle and partial wave analysis of this angle
dependence indicates presence of $l=2,4$ spherical harmonics along
with the angle independent $l=0$. We have found this extra presence
of D, G waves in angle dependence only when we use
$e^{-constant*area}$ form of $f$. But when $f$ is spherically
symmetric, i.e. old Gaussian form, then there appears no D, G waves.
Thus D, G waves must be a property of $e^{-constant*area}$. So
$e^{-constant*area}$ may couple an l=0 meson-meson system to l=2,
4,... etc. systems. This means two possibilities 1) l=0 meson-meson
system may have t-matrix and phase shifts to l=2, 4,... final state
meson-meson systems  and 2) l=0 may couple to l=2, 4,... states as
intermediate states in a polarization potential \cite{wong}, through
$e^{-constant*area}$.

\section*{Acknowledgments}

We are thankful to Higher Education Commission (HEC) of Pakistan for
there financial support through Grant No. 17-5-3(Ps3-056)
HEC/Sch/2006.

\begin{figure}
\begin{center}
\epsfig{file=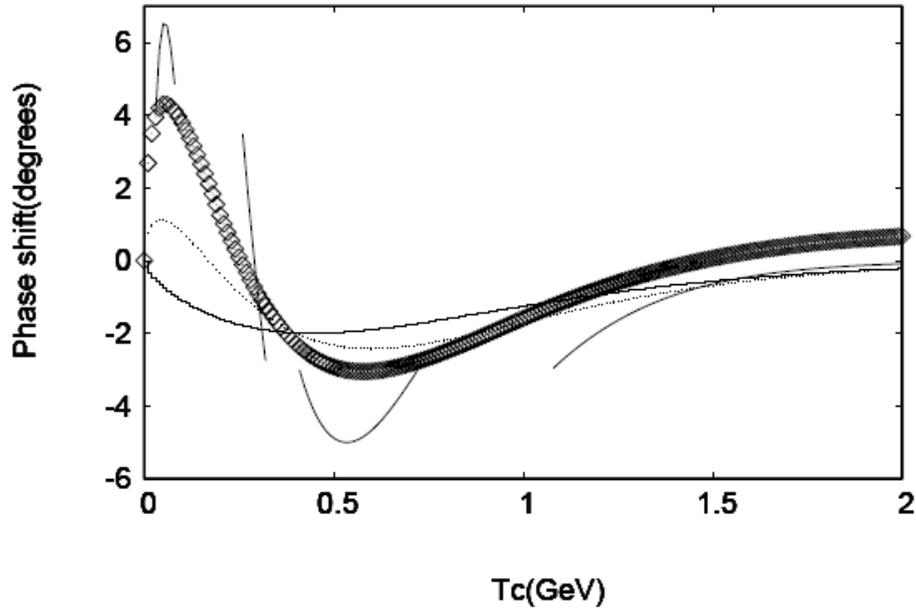,width=0.75\linewidth,clip=} \caption{Comparison of
different forms of $f$. Graph with lines only corresponds to
$k_{f}=0$. (Here data points are 0.01GeV apart but they are joined
for clarity. This graph is also shown separably in fig.\ref{graph7}.
Only four portions of the graph are shown here: upper peak
corresponds to actual data minus 28 and lower peak corresponds to
actual data plus 5.) Graph with dots corresponds to $S_{a}$ in $f$.
Graph with steps corresponds to gaussian form of $f$ for
$k_{f}=0.075$ as defined by eq.(\ref{e7}). Graph with points
corresponds to average of phase shifts for different values of angle
$\theta$ for $S$ in $f$ with $k_{f}=0.5$ defined by eq.(\ref{e6}).}
\label{graph2}
\end{center}
\end{figure}

\begin{figure}
\begin{center}
\epsfig{file=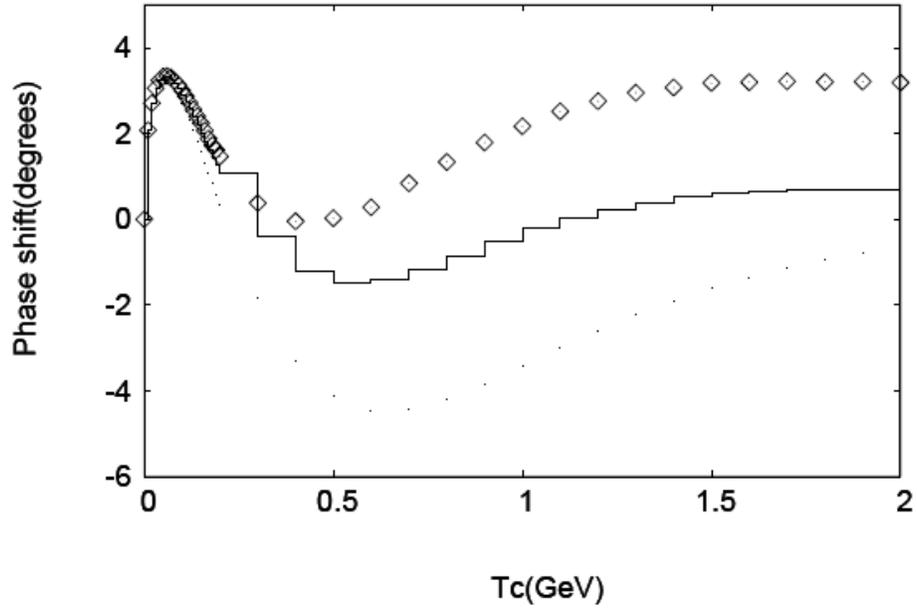,width=0.75\linewidth,clip=} \caption{Comparison of
phase shifts for different values of $\theta$
    using $S$ in $f$ with $k_{f}=0.57$. Choice of curve shapes is
    same as in fig. \ref{graph1}. But the results for $\theta=\pi/2$ are shown in fig. \ref{graph5} and not here.}
\label{graph3}
\end{center}
\end{figure}

\begin{figure}
\begin{center}
\epsfig{file=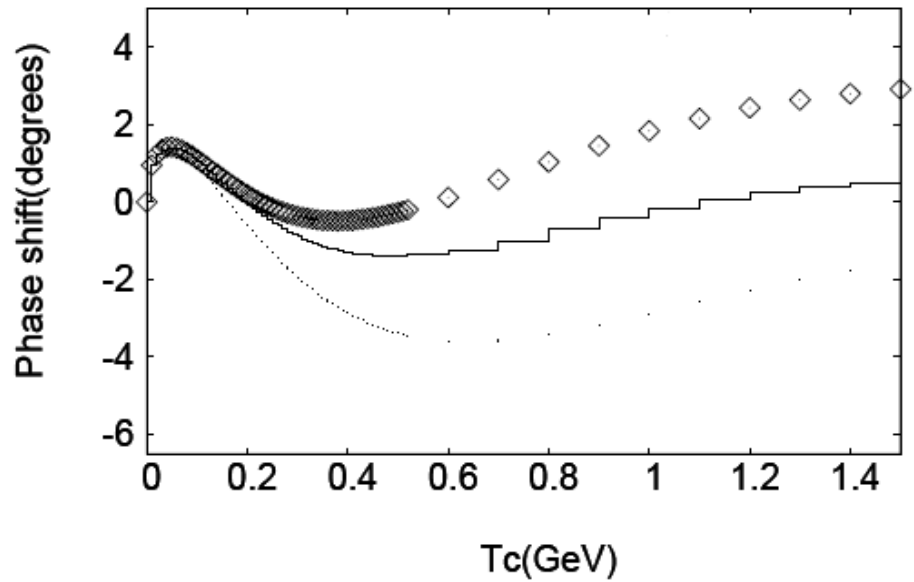,width=0.75\linewidth,clip=} \caption{Comparison of
phase shifts for different values of $\theta$
    using $S$ in $f$ with $k_{f}=0.8$. Choice of curve shapes is
    same as in fig. \ref{graph3}. The results for $\theta=\pi/2$ are shown in fig. \ref{graph6} below. }
\label{graph4}
\end{center}
\end{figure}

\begin{figure}
\begin{center}
\epsfig{file=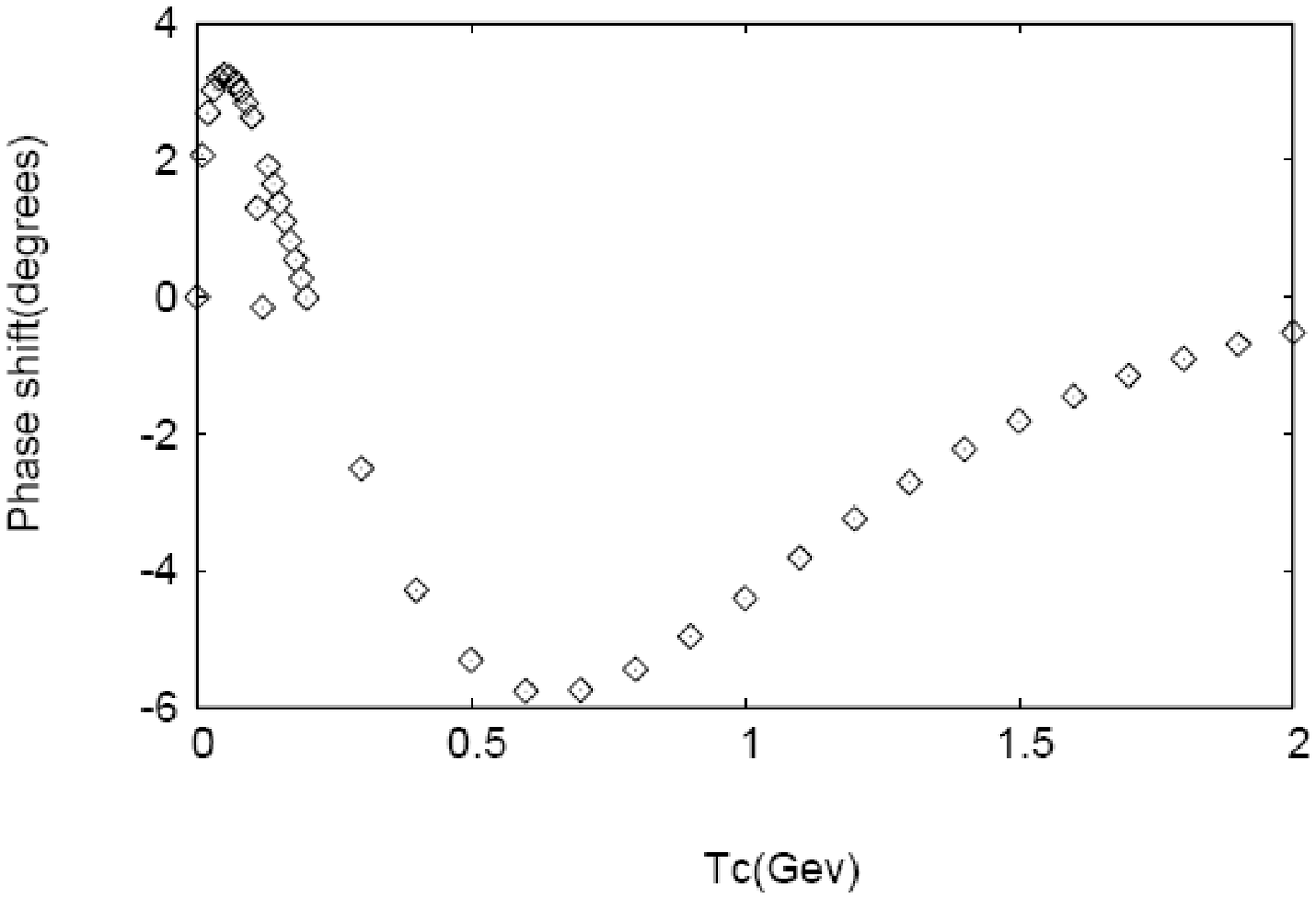,width=0.75\linewidth,clip=} \caption{Phase shifts
for $k_{f}=0.57$ at $\theta=\frac{\pi}{2}$ using $S$ in $f$.}
\label{graph5}
\end{center}
\end{figure}

\begin{figure}
\begin{center}
\epsfig{file=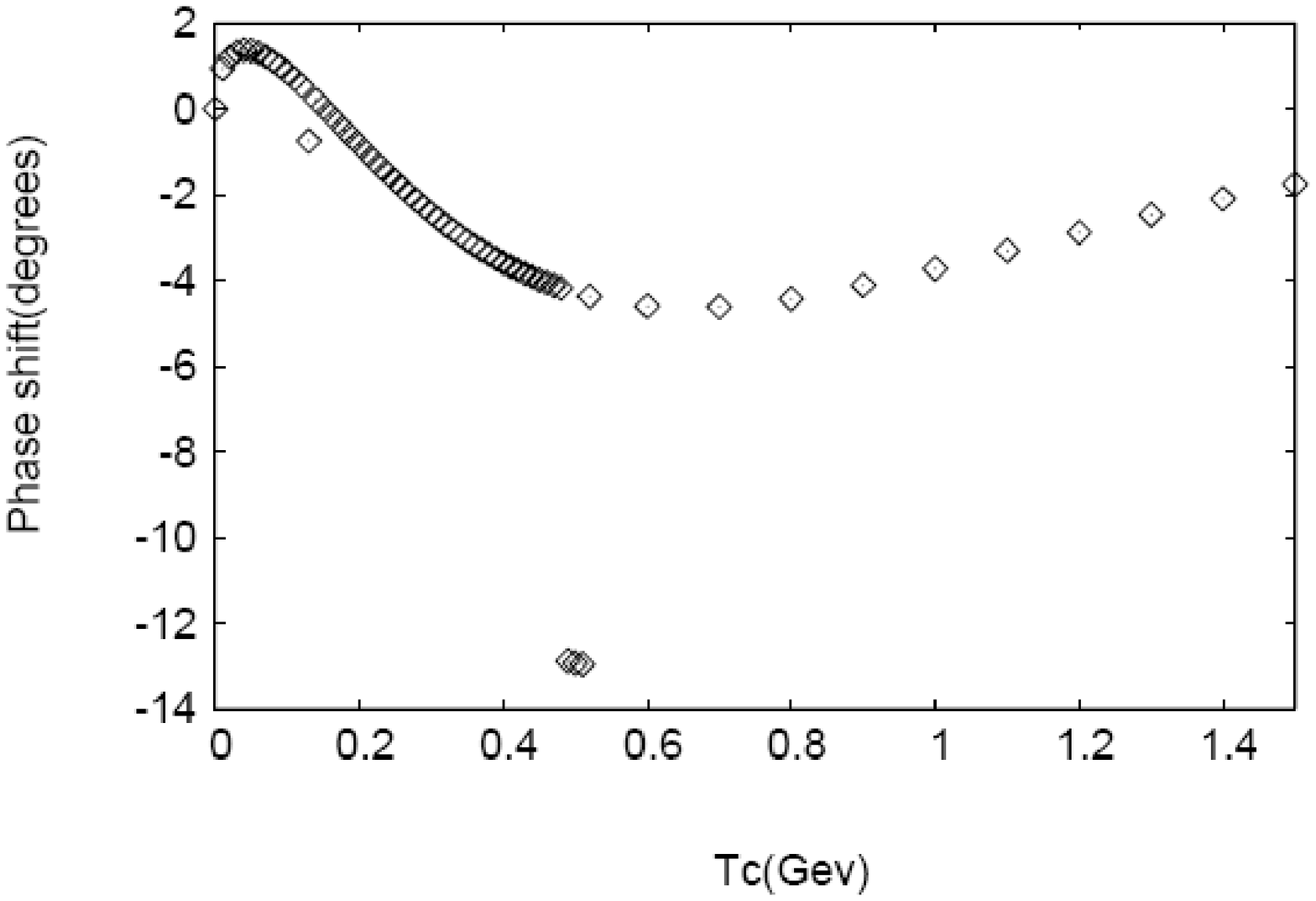,width=0.75\linewidth,clip=} \caption{Phase shifts
for $k_{f}=0.8$ at $\theta=\frac{\pi}{2}$ using $S$ in $f$.}
\label{graph6}
\end{center}
\end{figure}

\begin{figure}
\begin{center}
\epsfig{file=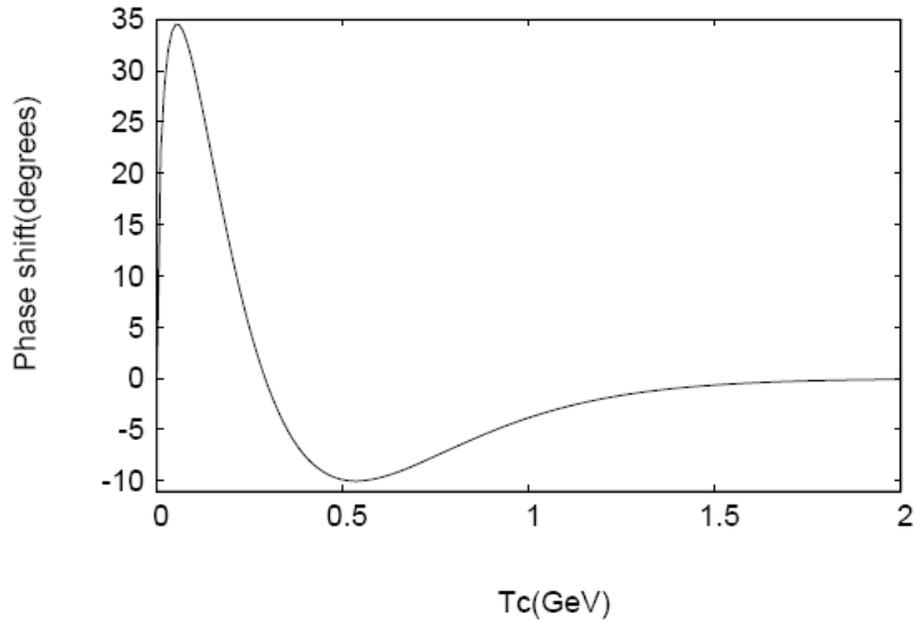,width=0.75\linewidth,clip=} \caption{Phase shifts
for $k_{f}=0$.} \label{graph7}
\end{center}
\end{figure}

\begin{figure}
\begin{center}
\epsfig{file=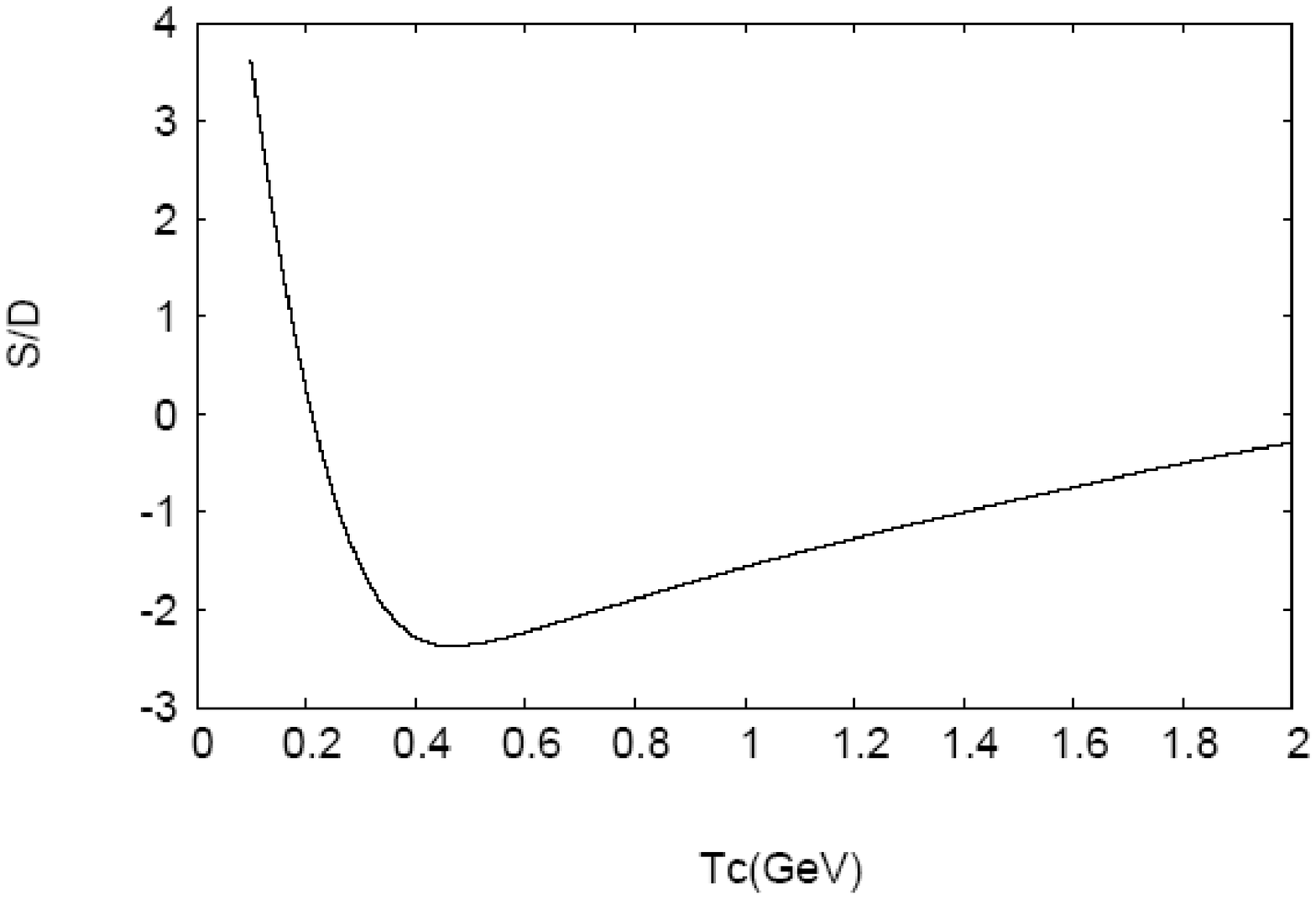,width=0.75\linewidth,clip=} \caption{S/D ratios
for different values of $T_{c}$.} \label{graph8}
\end{center}
\end{figure}

\begin{figure}
\begin{center}
\epsfig{file=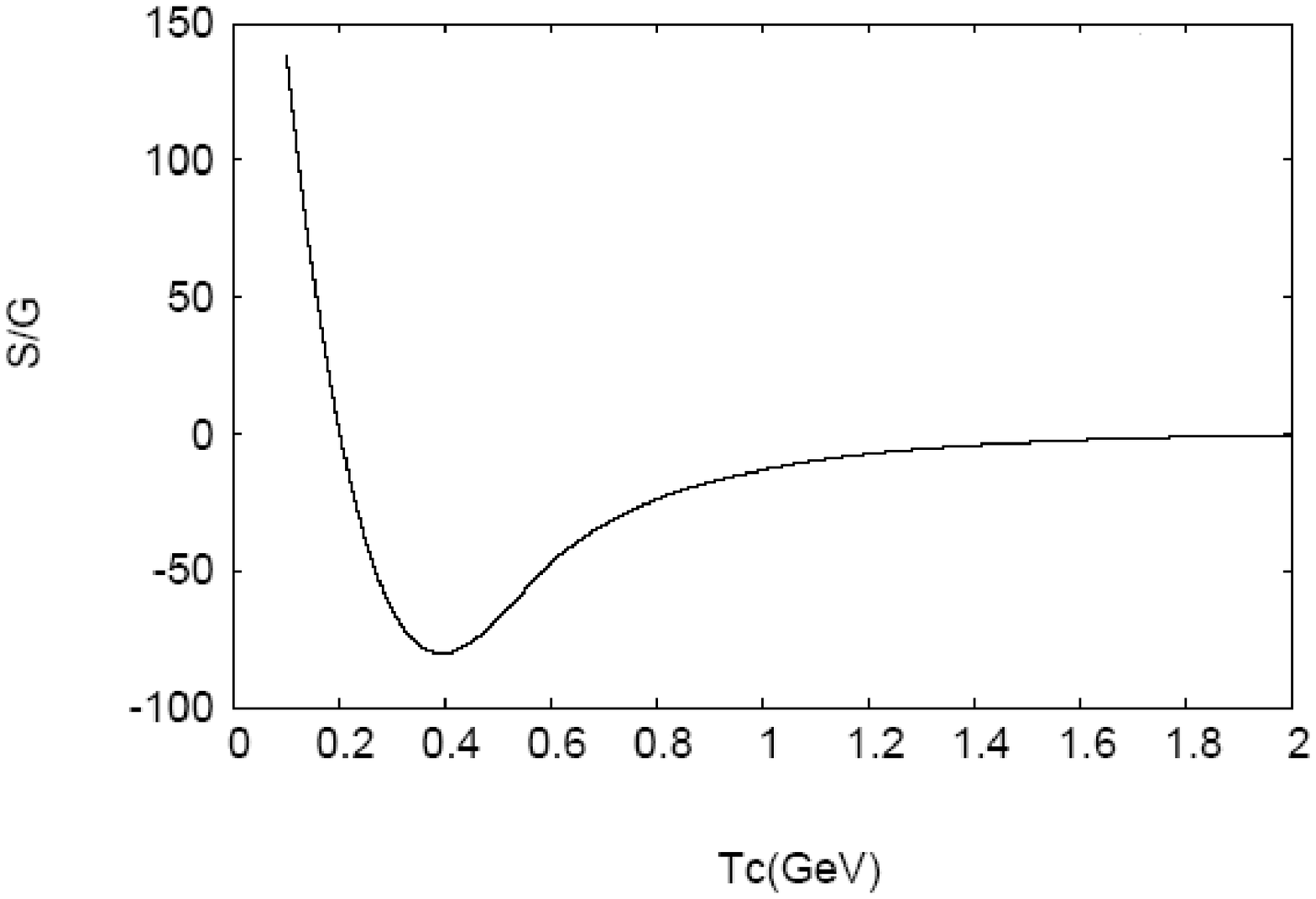,width=0.75\linewidth,clip=} \caption{S/G ratios
for different values of $T_{c}$.} \label{graph9}
\end{center}
\end{figure}

\begin{thebibliography}{999}
\bibitem{T. Barnes}
T. Barnes and E.S. Swanson, Phys.Rev. \textbf{D46} 131-159 (1992).
\bibitem{Green1}
A. M. Green, C. Michael, J.E. Paton, Nucl.Phys. \textbf{A554}
701-720 (1993).
\bibitem{B.Masud}
B. Masud, J.Paton, A.M. Green and G.Q. Liu, Nucl.phys. \textbf{A528}
477-512 (1991).
\bibitem{P. Pennanen}
A.M. Green and P. Pennanen, Phys.Rev. \textbf{C57} 3384-3391 (1998).
\bibitem{matsuoka} H. Matsuoka, D. Sivers, Phys.Rev. \textbf{D33} 1441 (1986).
\bibitem{green2}
A. M. Green, J. Koponen and P. Pennanen, Phys.Rev. \textbf{D61}
014014, (1999).
\bibitem{V. G. Bornyakov}
V. G. Bornyakov, P. Yu. Boyko, M. N. Chernodub and M. I. Polikarpov,
arXiv:hep-lat/0508006v1.
\bibitem{Hideo Suganuma}
Hideo Suganuma, Arata Yamamoto, Naoyuki Sakumichi, Toru T.
Takahashi, Hideaki Iida and Fumiko Okiharu, Mod. Phys. Lett.
\textbf{A23} 2331 (2008).
\bibitem{J. Vijande}
J. Vijande, A. Valcarce and J. M. Richard, Phys.Rev. \textbf{D76}
114013 (2007).
\bibitem{J. Weinstein}
J. Weinstein and N. Isgur, Phys.Rev.\textbf{D27} 588-599 (1983).
\bibitem{B. Silvestre-Brac}
B. Silvestre-Brac and C. Semay, Z.Phys. \textbf{C57} 273-282 (1993).
\bibitem{E.S. Swanson}
E.S. Swanson, Ann.Phys. (N.Y.) 220,73 (1992).
\bibitem{wheeler}
John Archibald Wheeler, Phys.Rev. \textbf{52} 1083 (1937).
\bibitem{ackleh}
E. S. Ackleh, T. Barnes and E. S. Swanson, Phys. Rev. \textbf{D54}
6811 (1996).
\bibitem{O. Morimatsu}
C. Alexandrou, T. Karapiperis, O. Morimatsu, Nucl.Phys.
\textbf{A518} 723-751 (1990).
\bibitem{masud}
B. Masud , Phys. Rev. \textbf{D50} 6783-6803 (1994).
\bibitem{P. Bicudo}
P. Bicudo and M. Cardoso, arXiv:1010.0281.
\bibitem{Fumiko Okiharu}
Fumiko Okiharu, Hideo Suganuma and Toru T. Takahashi, Phys.Rev.
\textbf{D72} 014505 (2005).
\bibitem{wong}
Cheuk-Yin Wong, Phys.Rev. \textbf{C69} 055202 (2004),
arXiv:hep-ph/0311088v2.
\bibitem{next}
M. Imran Jamil and Bilal Masud, in progress.
\bibitem{John Weinstein}
J. Weinstein and N. Isgur, Phys.Rev. \textbf{D41} 2236 (1990).
\bibitem{Isgur}
N. Isgur and J. Paton, Phys.Rev. \textbf{D31} 2910 (1985).
\bibitem{Teper}
M. Teper, Phys. Lett. \textbf{B397} 223 (1997).
\bibitem{pennanen}
Petrus Pennanen, Phys.Rev. \textbf{D55} 3958 (1997).
\bibitem{green}
S. Furui, A.M. Green and B. Masud, Nucl.Phys \textbf{A582} 682-696
(1995).
\bibitem{dawood}
D. Ahmad and B. Masud, in progress.
\end{thebibliography}
\end{document}